\documentclass[twocolumn,showpacs,preprintnumbers,amsmath,amssymb]{revtex4}

\usepackage{graphicx}
\usepackage{rotating}

\begin{document}

\draft

\title{Precision Measurements of Collective Oscillations in the BEC-BCS Crossover}

\author{A. Altmeyer,$^{1}$ S. Riedl,$^{1,2}$ C. Kohstall,$^{1}$ M.~J. Wright,$^{1}$ R. Geursen,$^{1}$ M. Bartenstein,$^{1}$ C. Chin,$^{3}$
J. Hecker Denschlag,$^{1}$ and R. Grimm$^{1,2}$}

\address{$^{1}$Inst.\ of Experimental Physics and Center for Quantum Physics, Univ.\ Innsbruck,
6020 Innsbruck, Austria\\$^{2}$Inst.\ for Quantum Optics and
Quantum Information, Acad.\ of Sciences, 6020 Innsbruck,
Austria\\$^{3}$James Franck Institute, Physics Department of the
University of Chicago, Chicago, IL 60637, USA}

\date{\today}

\pacs{34.50.-s, 05.30.Fk, 39.25.+k, 32.80.Pj}

\begin{abstract}
We report on precision measurements of the frequency of the radial
compression mode in a strongly interacting, optically trapped
Fermi gas of $^6$Li atoms. Our results allow for a test of
theoretical predictions for the equation of state in the BEC-BCS
crossover. We confirm recent quantum Monte-Carlo results and rule
out simple mean-field BCS theory. Our results show the long-sought
beyond-mean-field effects in the strongly interacting BEC regime.
\end{abstract}

\maketitle


Ultracold, strongly interacting Fermi gases
\cite{Ohara2002,Bourdel2003,Jochim2003,Bartenstein2004,Regal2004,Zwierlein2004,Bourdel2004,Kinast2004,
Bartenstein2004a,Kinast2004a,Chin2004a,Zwierlein2005,Partridge2005}
have attracted considerable attention over the past few years,
serving as unique model systems to create, control, and
investigate novel states of quantum matter. Experimentally, the
availability of such systems has opened up exciting possibilities
to study many-body quantum phenomena like molecular Bose-Einstein
condensation (BEC) \cite{Jochim2003} and the crossover from BEC to
a Bardeen-Cooper-Schrieffer (BCS) type superfluid
\cite{Bartenstein2004,Regal2004,Zwierlein2004,Bourdel2004,Kinast2004,
Bartenstein2004a,Chin2004a,Kinast2004a,Zwierlein2005,Partridge2005}.
These experiments may also lead to a better understanding of
strongly interacting quantum systems in different areas of
physics, ranging from high-T$_c$ superconductors to neutron stars
and the quark-gluon plasma.

A degenerate two-component Fermi gas undergoes the BEC-BCS
crossover \cite{Eagles1969}
when the $s$-wave scattering length $a$ is varied from positive to
negative values across a scattering resonance. In the crossover
region, where $a$ is comparable with or larger than the
interparticle spacing, the equation of state is governed by
many-body effects. Understanding the equation of state is a
fundamentally important challenge and constitutes a difficult task
for many-body quantum theories, even in the zero-temperature
limit. Mean-field BCS theory \cite{Eagles1969}
provides a reasonable interpolation between the well-understood
limits. More sophisticated crossover approaches \cite{Pieri2005}
yield quantitatively different results in certain regimes, none of
them however providing a complete description of the problem. The
most advanced theoretical results were obtained by numerical
calculations based on a quantum Monte-Carlo (QMC) approach
\cite{Astrakharchik2004}.

On the BEC side of the crossover, there is an interesting
competition in the equation of state between the strong
interactions in a Bose gas and the onset of fermionic behavior.
For a strongly interacting Bose gas,
one can expect 
quantum depletion to increase the average energy per particle. To
lowest order, this beyond-mean-field effect leads a correction to
the equation of state predicted by Lee, Huang, and Yang (LHY)
almost 50 years ago \cite{LHY}. Beyond mean-field effects are
expected to reduce the compressibility of a strongly interacting
Bose gas as compared to the weakly interacting case. However, when
approaching the resonance, fermionic behavior emerges and the
system loses its purely bosonic character, which increases the
compressibility of the strongly interacting gas. Mean-field BCS
theory does not contain beyond-mean field effects and the LHY
correction is absent there. However, the QMC results predict
beyond-mean-field effects to be visible on the BEC-side of the
crossover \cite{Astrakharchik2004}.

In this Letter, we report on precision measurements of the radial
compression mode 
in an optically trapped, strongly interacting Fermi gas of $^6$Li
atoms. The mode serves as a sensitive probe for the
compressibility and thus the equation of state of a superfluid gas
in the BEC-BCS crossover \cite{modes_theo,Astrakharchik2005}. We
reach a precision level that allows us to distinguish between the
predictions resulting from mean field BCS theory and QMC
calculations. Previous experiments on collective modes, performed
at Duke University \cite{Kinast2004,Kinast2004a} and at Innsbruck
University \cite{Bartenstein2004a}, showed frequency changes in
the BEC-BCS crossover in both the slow axial mode and the fast
radial compression mode of a cigar-shaped sample. The accuracy,
however, was insufficient for a conclusive test of the different
many-body theories in the strongly interacting regime.

We prepare a strongly interacting, degenerate gas of $^6$Li atoms
in the lowest two internal states as described in our previous
publications \cite{Bartenstein2004,Bartenstein2004a,Chin2004a}.
The broad Feshbach resonance centered at a magnetic field of
$B=834$\,G facilitates precise tuning of the scattering length $a$
\cite{Bartenstein2005}. Forced evaporative cooling is performed in
a 1030-nm near-infrared laser beam focussed to a waist of
54\,$\mu$m at $764$\,G. This results in a deeply degenerate cloud
of $N = 2.0(5)\times10^5$ atoms. By adiabatically increasing the
trap laser power after cooling, the sample is recompressed to
achieve nearly harmonic confinement. In the axial direction the
gas is magnetically confined in the curvature of the field used
for Feshbach tuning with an axial trap frequency of $\omega_z/2\pi
= 22.4$\,Hz at 834\,G. The experiments reported here are performed
at two different final values of the laser power of the
recompressed trap. At 135\,mW (540\,mW), the trap is 1.8$\mu$K
(7.3$\mu$K) deep and the radial trap frequency is $\omega_{r}
\approx 2\pi\times 290$\,Hz ($590$\,Hz). The Fermi energy of a
non-interacting cloud is calculated to $E_F = \hbar^2
k_F^2/2m=\hbar (3\omega_{r}^2\omega_z N)^{1/3} =k_B \times
500\,$nK ($800$\,nK); here $m$ is the mass of an atom and $k_B$ is
Boltzmann's constant.

Since our first measurements on collective excitation modes
\cite{Bartenstein2004a}, we have upgraded our apparatus with a
two-dimensional acousto-optical deflection system for the trapping
beam and a new imaging system along the trapping beam axis. These
two improvements provide us with full access to manipulate and
observe the radial motion.

\begin{figure}
\includegraphics[width=8cm]{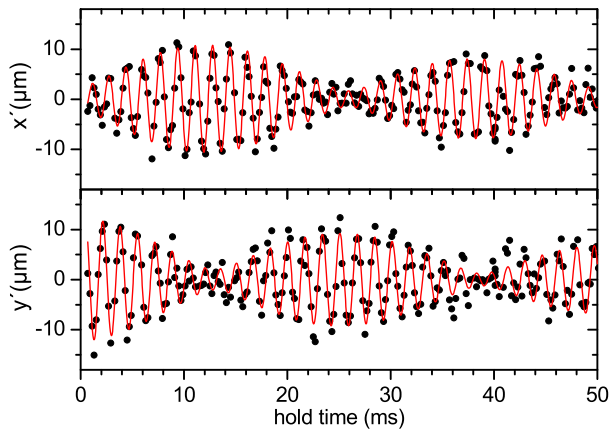}
\caption{Radial sloshing observed at a trap power of 540\,mW and $B = 735$\,G
($1/k_Fa=1.55$). The two-dimensional center-of-mass motion is represented in a
coordinate system ($x', y'$) rotated by 45$^{\circ}$ with respect to the
principal axes of the trap. The beat signal between the two sloshing eigenmodes
demonstrates the ellipticity of the trap with the two eigenfrequencies
$\omega_x/2\pi = 570$\,Hz and $\omega_y/2\pi = 608$\,Hz (ellipticity $\epsilon
= 0.066$).} \label{sloshing}.
\end{figure}

The trap beam profile is somewhat elliptic because of
imperfections and aberrations in the optical set up. To
simultaneously excite the two eigenmodes of the radial sloshing
motion, we initially displace the trapped sample into a direction
between the horizontal and vertical principal axes of the radial
potential. After a variable hold time, during which the cloud
oscillates freely, we turn off the optical trap. After a time of
flight of typically 4\,ms we take an absorption image of the
released cloud. The center-of-mass position of the cloud then
reflects its momentum at the instant of release. The experimental
results in Fig.~\ref{sloshing} demonstrate the sloshing with a
beat between the two eigenmodes. A careful analysis of such data
\cite{dr_alex} allows us to determine the eigenfrequencies
$\omega_x$ (horizontal sloshing) and $\omega_y$ (vertical
sloshing) to within a relative uncertainty of typically $2 \times
10^{-3}$. We finally derive the mean sloshing frequency
$\omega_\perp = \sqrt{\omega_x\omega_y}$ and the ellipticity
parameter $\epsilon=(\omega_y - \omega_x)/\omega_\perp$.

\begin{figure}
\includegraphics[width=8cm]{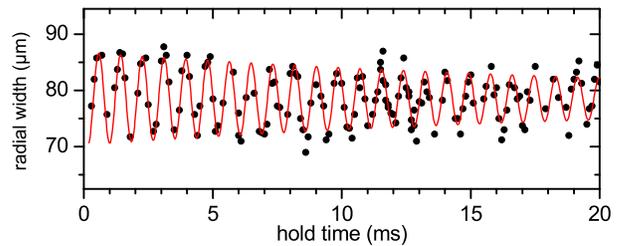}
\caption{Radial compression oscillation observed for the same conditions as the
sloshing mode data in Fig.~\ref{sloshing}. The radial width is determined by
averaging the horizontal and vertical Thomas-Fermi radii after expansion. Here
we obtain $\omega_c/2\pi = 1185$\,Hz.} \label{compression}
\end{figure}

To excite the radial compression oscillation we reduce the trap
light power for a short time interval of $\sim$$100\mu$s, inducing
an oscillation with a relative amplitude of typically $10\%$.
After a variable hold time the cloud is released from the trap.
From fits of two-dimensional Thomas-Fermi profiles to images of
the expanding cloud taken 4ms after release, we determine the mean
cloud radius. A typical set of measurements is shown in
Fig.~\ref{compression}. A fit of a damped harmonic oscillation to
such data yields the frequency $\omega_c$ and damping rate
$\gamma$ of the radial compression mode.

Our experiments are performed close to the limit of an elongated
harmonic trap potential with cylindrical symmetry. This elementary
case is of great general relevance for many quantum gas
experiments in optical and magnetic traps (see, e.g.,
\cite{Chevy2002}), and collective excitations are conveniently
normalized to the trap frequency $\omega_r$
\cite{modes_theo,Astrakharchik2005}. The compression mode
frequency can then be written as $\omega_c = f_c \omega_r$, where
$f_c$ is a dimensionless function of the interaction parameter
$1/k_Fa$ and is related to an effective polytropic index $\Gamma$
\cite{modes_theo,Astrakharchik2005} of the equation of state by
$\omega_c^2 = 2(\Gamma +1) \omega_r^2$.

In order to compare our experimental results with theory, we
consider the quantity $f_c$, i.e.\ the normalized compression mode
frequency of the ideal, cylindrically symmetric, elongated trap.
We find, that for our experimental conditions, $f_c$ is
approximated by the ratio $\omega_c/\omega_\perp$ of the measured
compression mode ($\omega_c$) and mean sloshing mode
($\omega_\perp$) frequencies to better than one percent. On the
desired accuracy level of $10^{-3}$, however, two small effects
have to be taken into account: the residual trap ellipticity and
the anharmonicity of the radial potential in combination with the
spatial extension of the trapped sample. We thus introduce two
small corresponding corrections, expressing $f_c$ in the form $f_c
= (1-\kappa \epsilon^2 + b\alpha) \, \omega_c/\omega_\perp$.

For the ellipticity correction $\kappa \epsilon^2$, a
straightforward solution of the hydrodynamic eigenfrequency
equation \cite{dr_alex} yields $\kappa = (2+\Gamma)/4\Gamma$,
where $\Gamma$ can be approximated by
$\Gamma=(\omega_c/\omega_\perp)^2/2-1$. For the anharmonicity
correction, the parameter $\alpha = \frac{1}{2}m \omega_\perp^2
r_{\rm rms}^2 /U_0$ relates the potential energy associated with
the root-mean-square radius $r_{\rm rms}$ \cite{rms} of the
trapped cloud to the trap depth. The coefficient $b$ results from
the differential anharmonicity shifts in the compression and
sloshing modes and can be calculated according to
\cite{anharmonic,kinastphd,dr_alex}. We obtain \cite{dr_alex}
$b=0.167$ and $0.280$ in the limits of BEC and unitarity,
respectively.

\begin{table}
\caption{Experimental data on radial collective modes in the
BEC-BCS crossover. The data in the upper seven (lower eight) rows
refer to the sets of measurements taken in the shallower (deeper)
trap with $U_0 = 1.8\mu$K and $E_F = 500$\,nK ($U_0 = 7.3\mu$K and
$E_F = 800$\,nK). The values in parentheses indicate $1\sigma$ fit
uncertainties of individual measurements. Note that a systematic
scaling uncertainty of $\sim$$4\%$ for $1/k_Fa$ results from the
uncertainty in the atom number $N = 2.0(5)\times10^5$.}
\begin{tabular}{|cc|cc|cc|cc|}\hline
      &           & \multicolumn{2}{c|}{sloshing} & \multicolumn{2}{c|}{compression} &  \multicolumn{2}{c|}{corr.}\\
  $B$ & $1/k_Fa$  &  $\omega_\perp/2\pi$ & $\epsilon$ & $\omega_c/2\pi$  & $\gamma/\omega_\perp$ & $\kappa\epsilon^2$ & $b\alpha$\\
  (G) &        &  (Hz)                 &         & (Hz) & & \multicolumn{2}{c|}{($10^{-4}$)} \\
  \hline
  727.8  & 2.21  & 292.7(5) & 0.083(3) & 596.3(6) & 0.007(2) & 48 & 20 \\
  735.1  & 1.96  & 298.6(5) & 0.091(3) & 602.8(8) & 0.008(3) & 60 & 26 \\
  742.5  & 1.75  & 294.5(5) & 0.067(3) & 593.2(7) & 0.005(2) & 33 & 28 \\
  749.8  & 1.55  & 296.3(4) & 0.073(3) & 599.0(7) & 0.006(2) & 38 & 28 \\
  760.9  & 1.27  & 296.0(4) & 0.088(2) & 592.3(7) & 0.009(2) & 58 & 24 \\
  771.9  & 1.03  & 293.6(7) & 0.074(5) & 586.2(8) & 0.007(3) & 41 & 27 \\
  834.1  & 0     & 287.5(7) & 0.073(5) & 519.4(9) & 0.014(3) & 55 & 94 \\ \hline
  757.2  & 1.07  & 605.0(9) & 0.065(3) & 1210.9(12) & 0.010(2) & 32 & 13 \\
  768.2  & 0.87  & 592.5(7) & 0.069(2) & 1186.6(12) & 0.012(2) & 36 & 16 \\
  775.6  & 0.75  & 590.2(4) & 0.060(1) & 1170.2(21) & 0.007(4) & 28 & 14 \\
  782.2  & 0.64  & 604.8(9) & 0.061(3) & 1187.1(16) & 0.006(3) & 29 & 16 \\
  801.3  & 0.38  & 586.8(7) & 0.063(2) & 1135.2(12) & 0.010(2) & 33 & 24 \\
  812.3  & 0.24  & 586.5(7) & 0.058(2) & 1106.9(16) & 0.014(3) & 30 & 33 \\
  834.1  & 0     & 596.3(9) & 0.070(3) & 1089.0(12) & 0.010(2) & 48 & 40 \\
  849.1  & -0.14 & 583.2(7) & 0.052(2) & 1046.7(37) & 0.007(2) & 29 & 47 \\ \hline
\end{tabular}
\label{Table1}
\end{table}

Our measurements on the sloshing and compression modes are
summarized in Table \ref{Table1}, including the two small
corrections. For the data in the strongly interacting BEC regime
($1/k_Fa\gtrsim 1$) we used the weaker trap with
$\omega_\perp/2\pi \approx 290$\,Hz to minimize unwanted heating
by inelastic collisions. Closer to resonance ($1/k_Fa \lesssim 1$)
inelastic processes are strongly suppressed, but the increasing
cloud size introduces larger anharmonicity shifts. Here we chose
the deeper trap with $\omega_\perp/2\pi \approx 590$\,Hz. On the
BCS side of the resonance we observed increased damping as a
precursor of the breakdown of hydrodynamics
\cite{Bartenstein2004a,Kinast2004a}. We thus restricted our
measurements to magnetic fields below 850\,G to ensure low damping
rates ($\gamma/\omega_\perp < 0.01$) and superfluid hydrodynamics.

At a given magnetic field, a set of measurements on the sloshing
and compression modes 
typically takes a few hours. 
To minimize uncertainties from slow drifts
and day-to-day variations we always took the sloshing mode
reference measurement right before or after the compression mode
data. By repeating measurements under identical settings we found
a typical remaining fractional uncertainty for the normalized
compression mode frequencies of $5\times10^{-3}$, which is about
2-3 times larger than the fit errors of individual measurements.

\begin{figure}[t]
\includegraphics[width=8cm]{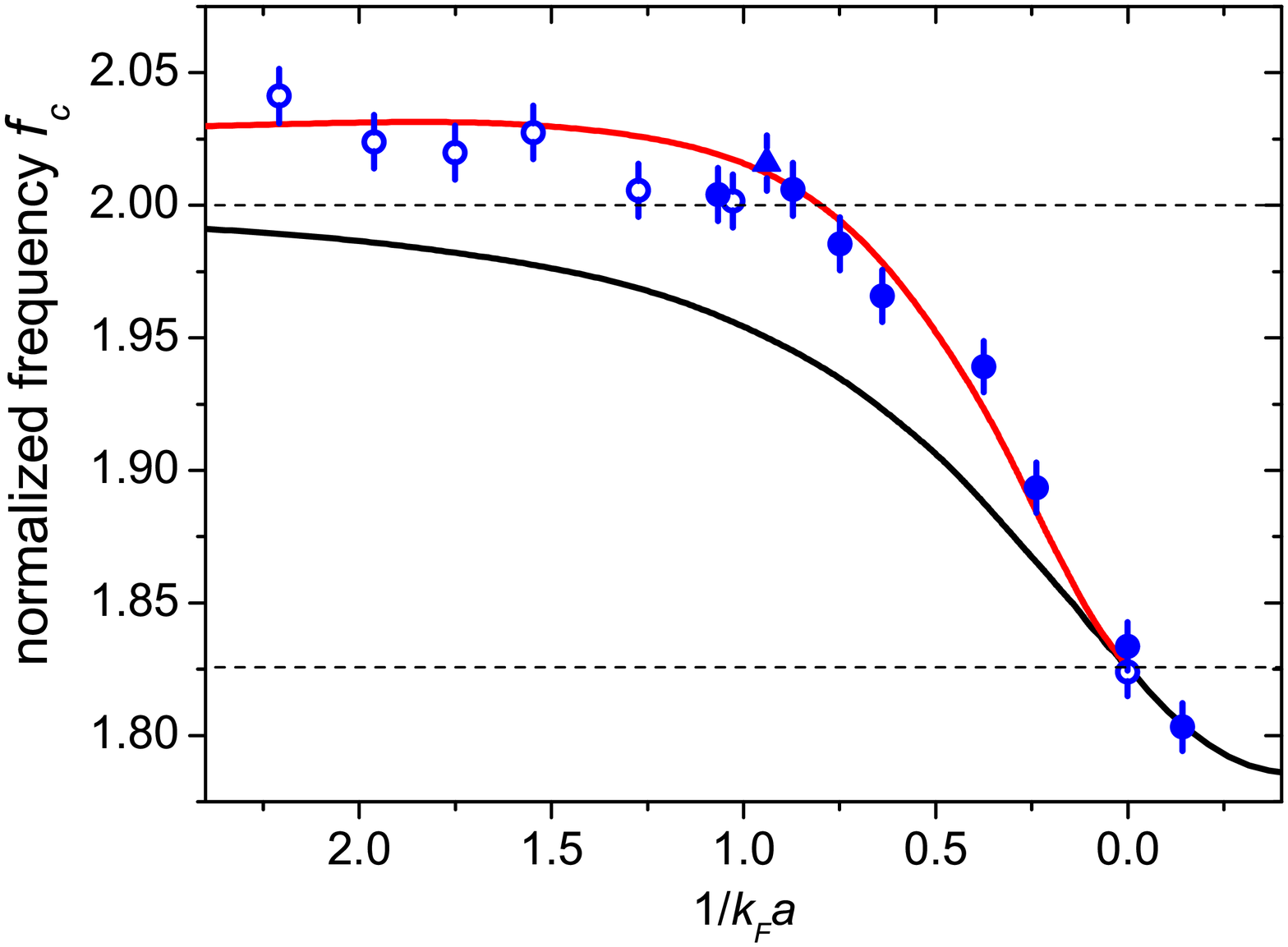}
\vspace{-3mm} \caption{Normalized compression mode frequency $f_c$
versus interaction parameter $1/k_Fa$. The experimental data
include the small corrections for trap ellipticity and
anharmonicity and can thus be directly compared to theory in the
limit of an elongated harmonic trap with cylindrical symmetry. The
open and filled circles refer to the measurements listed in
Table~\ref{Table1} for $\omega_\perp/2\pi \approx 290$\,Hz and
590\,Hz, respectively. The error bars indicate the typical scatter
of the data points. The filled triangle shows a zero-temperature
extra\-polation of the measurements displayed in
Fig.~\ref{finiteT}. The theory curves refer to mean-field BCS
theory (lower curve) and QMC calculations (upper curve) and
correspond to the data presented in Ref.~\cite{Astrakharchik2005}.
The horizontal dashed lines indicate the values for the BEC limit
($f_c = 2$) and the unitarity limit ($f_c=\sqrt{10/3}=1.826$).}
\label{crossover}
\end{figure}

In Fig.~\ref{crossover} we show our final results on the
normalized compression mode frequency in the BEC-BCS crossover.
The two theory curves \cite{Astrakharchik2005} correspond to the
equation of state from mean-field BCS theory (lower curve) and the
one from quantum Monte-Carlo calculations (upper curve). Our data
confirm the quantum Monte-Carlo predictions and rule out the
mean-field BCS theory. In the strongly interacting BEC regime
($1/k_Fa\gtrsim 1$) our data are well above the value of $2$. This
highlights the presence of the long-sought beyond-mean-field
effects \cite{LHY} in collective modes of a strongly interacting
gas \cite{Pitaevskii1998,modes_theo}.

\begin{figure}[t]
\includegraphics[width=6.3cm]{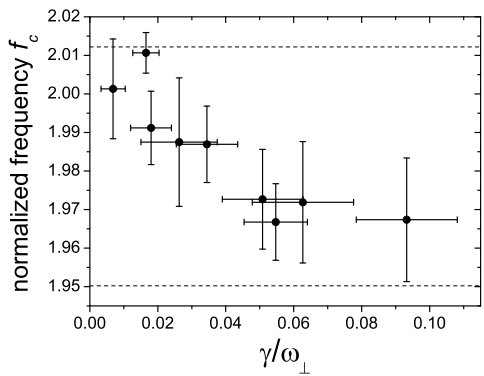}
\vspace{-2mm}
\caption{Normalized compression mode frequency $f_c$
versus damping rate for $1/k_Fa = 0.94$ ($U_0 = 7.3\mu$K). The
error bars represent $1\sigma$ fit uncertainties. The dashed lines
indicate the zero-temperature values predicted by QMC calculations
(upper line) and mean-field BCS theory (lower line).}
\label{finiteT}
\end{figure}

We finally address the question how non-zero temperatures
influence the compression mode frequency. At unitarity, a recent
experiment \cite{Kinast2005} has found small frequency upshifts
with temperature. For a BEC, however, theory \cite{Giorgini2000}
predicts temperature-induced down-shifts, which compete with the
up-shifts from beyond-mean-field effects. We have performed a set
of measurements on temperature shifts in the strongly interacting
BEC regime ($1/k_Fa=0.94$). Before exciting the collective
oscillation, the evaporatively cooled gas was kept in the
recompressed trap for a variable hold time of up to 1.5\,s. During
this time residual heating by inelastic processes slowly increased
the temperature, which we observed as a substantial increase of
damping with time. The damping rate $\gamma$ thus serves us as a
very sensitive, but uncalibrated thermometer
\cite{Kinast2004,Kinast2005}. Fig.~\ref{finiteT}, where we plot
the normalized compression mode frequency versus damping rate,
clearly shows a temperature-induced down-shift. We note that
previous measurements in the strongly interacting BEC regime
\cite{Bartenstein2004a,Kinast2004a} were performed at relatively
large damping rates in the range between 0.05 and 0.1, where
frequency down-shifts are significant.

With our new knowledge on systematic frequency shifts in
collective mode measurements, let us comment on the previous
experiments performed in Innsbruck \cite{Bartenstein2004a} and at
Duke University \cite{Kinast2004,Kinast2004a}. We have reanalyzed
our old data on the radial compression mode and identified a
previously undetermined ellipticity of $\epsilon \approx 0.2$ as
the main problem in our data interpretation
\cite{reinterpretation}. The fact that we had normalized the
compression mode frequency to the vertical trap frequency
($\omega_c/\omega_y$) led to a substantial down shift in the
hydrodynamic regime, but not in the collisionless regime. We
furthermore believe that significant temperature shifts were
present in the previous collective mode experiments. In particular
for the strongly interacting BEC regime, temperature shifts in our
old data on the axial mode \cite{Bartenstein2004a} and the Duke
data on the radial mode \cite{Kinast2004a} provide a plausible
explanation for these measurements being closer to the predictions
of mean-field BCS theory than to the more advanced QMC results.

In conclusion, our work shows that collective modes allow for
precision tests of many-body theories in strongly interacting
quantum gases. In future experiments, the observation of
collective oscillation modes will serve as a powerful tool to
investigate strongly interacting superfluids in a more general
context, e.g.\ in mixtures of fermionic quantum gases.

We warmly thank S.\ Stringari for stimulating our interest in
collective modes and for many useful discussions. We thank G.\
Astrakharchik for providing us with the theoretical data for
Fig.~\ref{crossover}, and R.\ Danielian for assistance in the
experiments. We acknowledge support by the Austrian Science Fund
(FWF) within SFB 15 (project part 21). S.R.\ is supported within
the Doktorandenprogramm of the Austrian Academy of Sciences. C.C.\
acknowledges travel support from the NSF-MRSEC program under
DMR-0213745.


\end{document}